\documentclass[prl,twocolumn,showpacs]{revtex4}

\usepackage{graphicx}
\usepackage{bm}

\def\w{\omega}

\begin{document}

\title{Renormalization Group Approach to  Spectral Properties of
  the Two-Channel Anderson Impurity Model} 

\smallskip

\author{Frithjof B.~Anders}
\affiliation{Department of Physics, Universit\"at Bremen,
                  P.O. Box 330 440, D-28334 Bremen, Germany}

\date{7-07-2004}

\begin{abstract}
The impurity Green function and dynamical susceptibilties for the
two-channel Anderson impurity model are calculated. An exact
expression for the self-energy of the impurity Green function is
derived. The imaginary part of the self-energy scales as
$\sqrt{|\w/T_K|}$ for $T\to 0$ serving as a hallmark for non-Fermi
behavior. The many-body 
resonance is pinned to a universal value $1/(2\pi\Delta)$ at
$\w=0$. Its shape becomes increasingly more symmetric for the
Kondo-regimes of the model. The  dynamical 
susceptibilities are governed by  two energy scales $T_K$
and $T_h$  and approach a constant value for $\w\to
0$, 
whereas relation $\chi''(\w)\propto \w$ holds for
the single channel model.

\end{abstract}

\maketitle

\def\e{\varepsilon}
\def\as{\alpha\sigma}
\def\ket#1{|#1\rangle}
\def\bra#1{\langle #1|}
\def\green#1#2{\ll \!\! #1| #2 \!\! \gg}
\def\sus{susceptibility}

\paragraph{Introduction:}
In the last decade, the paradigm of a Fermi-liquid ground state in Heavy
Fermion (HF) materials \cite{Grewe91}, which originally gave
these mainly  Ce and Uranium base compounds their name, has become
strongly questioned. In addition to magnetic and superconducting phase
transitions, many of these alloys show deviations  from  the
Fermi-liquid behavior in the normal phase. Typically, logarithmic
behavior and power laws with anomalous exponents  have been reported
in the low-temperature thermodynamic and transport properties
\cite{Stewart01}.  These unusual properties are attributed to  charge
and spin fluctuations on the partially filled $f$-shells. Certain
ballistic metallic point contacts, two-level systems\cite{CoxZawa98}
and   coupled quantum-dot\cite{ALS2004} have been argued to display
similarly unusual properties.

While the $4f$-shell of $Ce$ ions is usually occupied by one electron,
the configuration of the $5f$-shell of U ions  fluctuates between
degenerate $5f^2$ and $5f^3$ crystal field states. Hund's rules
coupling and crystal field splitting in a cubic symmetry yield a
Kramer's degenerate $\Gamma_6$ ground state $\ket{\sigma}$ for the
$5f^3$ and a quadrupolar $5f^2$-$\Gamma_3$ 
doublet $\ket{\alpha}$. On symmetry grounds, these states only
hybridized with $\Gamma_8$ conduction electron labelled by a spin
$\sigma$ and channel index $\alpha$\cite{Cox87}. 
The Hamiltonian of this two-channel Anderson impurity model (TC-SIAM)
is  given by  
\begin{eqnarray}
\label{eqn:tcsiam-h}
 {\cal H} & =&  \sum_{k\alpha\sigma} \e_{k\as} c^\dagger_{k\as} c_{k\as}
+ \sum_{\sigma}(E_\sigma -E_\alpha) X_{\sigma,\sigma}
+ h_{s} S_z
\nonumber \\
&& 
+ h_{c} \tau_z
+ \sum_{k\as} V_\alpha
\left(
X_{\sigma,-\alpha}
c_{k\as}
+c^\dagger_{k\as} 
X_{-\alpha,\sigma}
\right)
\end{eqnarray}
where the operator $ c^\dagger_{k\as}$ creates an electron with energy
$\e_{k\as}$, momentum $k$, quadrupolar channel $\alpha=\pm$ and spin
$\sigma$. $X_{m,m'} =\ket{m}\bra{m'}$ is the usual Hubbard operator
which destroys the ionic state $\ket{m'}$ and creates the  ionic state
$\ket{m}$. The energy $E_\sigma$ is assigned to the magnetic doublet,
the energy $E_\alpha$ to the quadrupolar doublet containing one
electron less than $\ket{\sigma}$. The last term describes the
hybridization between the $f$-shell and the conduction electron host.
The external magnetic and channel fields, $h_s$ and $h_c$, 
couple to the local spin or channel flavor operator, $S_z= \sum_{\sigma}
\sigma X_{\sigma\sigma}/2$, and $\tau_z =\sum_{\alpha} \alpha
X_{\alpha\alpha}/2$ and are used to calculate the local spin- or
channel susceptibility  $\chi_{s} = \frac{\langle S_z\rangle}{h_s}$ or  
$\chi_{c} =\frac{ \langle \tau_z\rangle}{h_c}$, respectively, for
$h_s,h_c\to 0$. 
This model was proposed in the context of UBe$_{13}$ by
Cox\cite{Cox87,CoxZawa98} for a novel and unusual Kondo effect: the 
non-magnetic $5f^2$-$\Gamma_3$   ground state undergoes a quadrupolar
charge  rather than a magnetic  Kondo-screening. Using a 
Monte-Carlo approach\cite{SchillerAndCox98}, it was shown that (i) the
magnetic and quadrupolar moments are screened for $T\to 0$, (ii) this screening
is described by two different energy scales $T_h\ge T_l\to T_K$, (iii)
both  channel and spin susceptibility   behave as $\chi_{s/c} \propto
\ln(T)$ in the universality regime $T\ll T_l$. Recently, this was
confirmed by an exact Bethe {\em ansatz} solution\cite{BolechAndrei02}
of the model (\ref{eqn:tcsiam-h}). 
It was also possible to calculate the threshold exponents of the
resolvent $I_M(z) = \bra{M,0}1/(z-\hat H)\ket{M,0} \approx I_0
(z-E_g)^{-\alpha_M}$ of the time evolution of the local levels
$M=\sigma,\alpha$ relative to the Fermi-see of the free electron
gas\cite{JohnnaessonBolechAndrei03}; $E_g$ is the threshold energy. In
contrary to the  non-Crossing Approximation
(NCA)\cite{Grewe83,Kuramoto83,Cox87,CoxZawa98} predicting $\alpha_M=1/2$, the true 
exponents are dependent on the occupation of $n_c=\sum_\sigma \langle
X_{\sigma\sigma} \rangle $. Since  the Green function $G_{\as}(z)= \ll
X_{\alpha,\sigma} | X_{\sigma,\alpha}\gg$  is obtained by a
convolution of $I_{\sigma}(z)$ and $I_{\alpha}(z)$ in NCA
\cite{Cox87}, its spectral function would be given by $\rho_{\as}(\w)
= \rho_0^\pm |\w|^{1-\alpha_\sigma-\alpha_\alpha}
B(1-\alpha_\sigma,1-\alpha_\alpha) \propto \rho_0^\pm
|\w|^{-\frac{1}{2}n_c(n_c-1)} $ for $|\w|\to 0$ using the exact
threshold exponents\cite{CostiWoelfe94}; $B$ is the beta function.
Hence, the spectral function should diverge for $n_c\not =0,1$,
indicating that  vertex corrections are
as important as the exact exponents $\alpha_M$ for obtaining the correct
spectral function from $I_M(z)$\cite{Anders95}.

%
In this Letter,
the first accurate solution for the
dynamical properties of the  model using Wilson's numerical
renormalization-group method (NRG)\cite{Wilson75}
is presented. It is shown that  (i) the scaling 
behavior of $\rho_{\as}(\w)$ is given by $(1-a_\pm \sqrt{|\w|/T_K})$
with a very weak $n_c$ dependence of the constants $a_\pm$, 
(ii) the many-body resonance becomes increasingly more symmetric in
the Kondo regime in contrast to the NCA result, (iii)  the total
self-energy consists of a  resonant level term and  a local non-Fermi
liquid   self-energy with vanishing imaginary part at $\w=0$, (iv) the
dynamical spin and channel susceptibility $\chi''(\w)$ approach a
constant for $|\w|<T_K$ and $T\ll T_K$, corresponding to
logarithmically divergent $\chi(T)$ for $T\to 0$.

The Hamiltonian (\ref{eqn:tcsiam-h}) is solved using the
NRG\cite{Wilson75}. The core of the NRG approach is a logarithmic energy
discretization of the conduction band around the Fermi
energy, $\w_n^\pm = \pm D\lambda^{-n}$ and $\Lambda>1$,
and a unitary transformation of the base $c_{\w_n^\pm\as}$ onto a base
such that the Hamiltonian becomes tridiagonal. The first operator
$f_{0\as}$ is defined as $f_{0\as}=\frac{1}{2}\sum_{n}
(c_{\w_n^+\as}+c_{\w_n^-\as})$ \cite{Wilson75}. Only $f_{0\as}$
couples directly to the impurity degrees of freedom. 
Eqn.~(\ref{eqn:tcsiam-h}) is recasted
as a double limit of a sequence of dimensionless NRG
Hamiltonians:
\begin{equation}
{\cal H} = \lim_{\Lambda \rightarrow 1^+}
           \lim_{N \rightarrow \infty}
           \left\{
                 D_{\Lambda} \Lambda^{-(N-1)/2} {\cal H}_{N}
           \right\} ,
\end{equation}
with $D_{\Lambda}$ equal to $D(1 + \Lambda^{-1})/2$, and
\begin{eqnarray}
{\cal H}_N &=& \Lambda^{\frac{N-1}{2}}
               \left [
                       \frac{E_{\as}}{D_{\Lambda}}
                            \sum_{\sigma} X_{\sigma,\sigma}
+
               \sum_{\sigma\alpha}
               \left \{
                      \tilde{V}_\alpha f^{\dagger}_{\alpha \sigma 0}
                      X_{-\alpha\sigma} 
                      + {\rm H.c.}
               \right \}
              \right. 
\nonumber \\
&+&
               \left.
                     \sum_{n = 0}^{N-1}
                     \sum_{\alpha \sigma} \Lambda^{-\frac{n}{2}}
                     \xi_{n \alpha}
                     \left \{
                            f^{\dagger}_{\alpha \sigma n+1}
                            f_{\alpha \sigma n}
                            + {\rm H.c.}
                     \right \}
               \right ] .
\nonumber
\end{eqnarray}
$\tilde{V}_{\alpha}$ is  related to the hybridization width
$D_{\Lambda}\tilde{V}_{\alpha} = V_\alpha$, $\Delta E=E_{\as}
=E_\sigma-E_\alpha$,  while the pre-factor $\Lambda^{(N-1)/2}$
guarantees that the low-lying excitations of ${\cal H}_N$ are of order
one for all $N$\cite{Wilson75}. 
In the absence of a symmetry breaking channel or magnetic field, the
Hamiltonian has three fixed points: the free orbital fixed point
$H_{FO}^*$ with $ \Delta E=0,V=0$, the two local moment fixed points
$H_{LM}^*$, $|\Delta E| \to \infty, V=0$  \cite{KrishWilWilson80ab}  and an
intermediate coupling or two-channel fixed point $H_{tc}^*$ governing
the universality regime of the model for $T\to 0$.  $H_{FO}^*$ is
instable with respect to the two relevant operators 
$X_{\sigma\sigma}$ and the hybridization. the local moment fixed
point is  instable with respect to the Kondo interaction $H_{K}$, which is
either of  spin type when $\Delta E<0$, or of channel type when $\Delta E>0$.
There are two
cross-over energy scales in the 
problem: the high temperature scale  $T_h$,  associated with the
screening of the upper doublet, and  the low temperature scale
$T_l=T_K$, the Kondo temperature,  associated with the
screening of the lower doublet. The latter temperature is exponentially
dependent on $\Delta/|E_{\as}|$, and the Bethe {\em ansatz}
\cite{BolechAndrei02} predicts 
$T_K = \frac{4\Delta}{\pi^2} \exp[-\frac{\pi |\Delta E|}{2\Delta }]$
for the wide band limit.
Introducing the  hybridization function  $\Gamma_{\as}(z)= \sum_k
V_\alpha|^2/(z-\e_{k\as})
$, 
 the imaginary part at $\w=-i\delta$ is denoted by $\Delta =\Delta_{\as} = \Im m \Gamma_{\as}(-i\delta)$.

\begin{figure}[t]
  \centering
  \includegraphics[width=89mm]{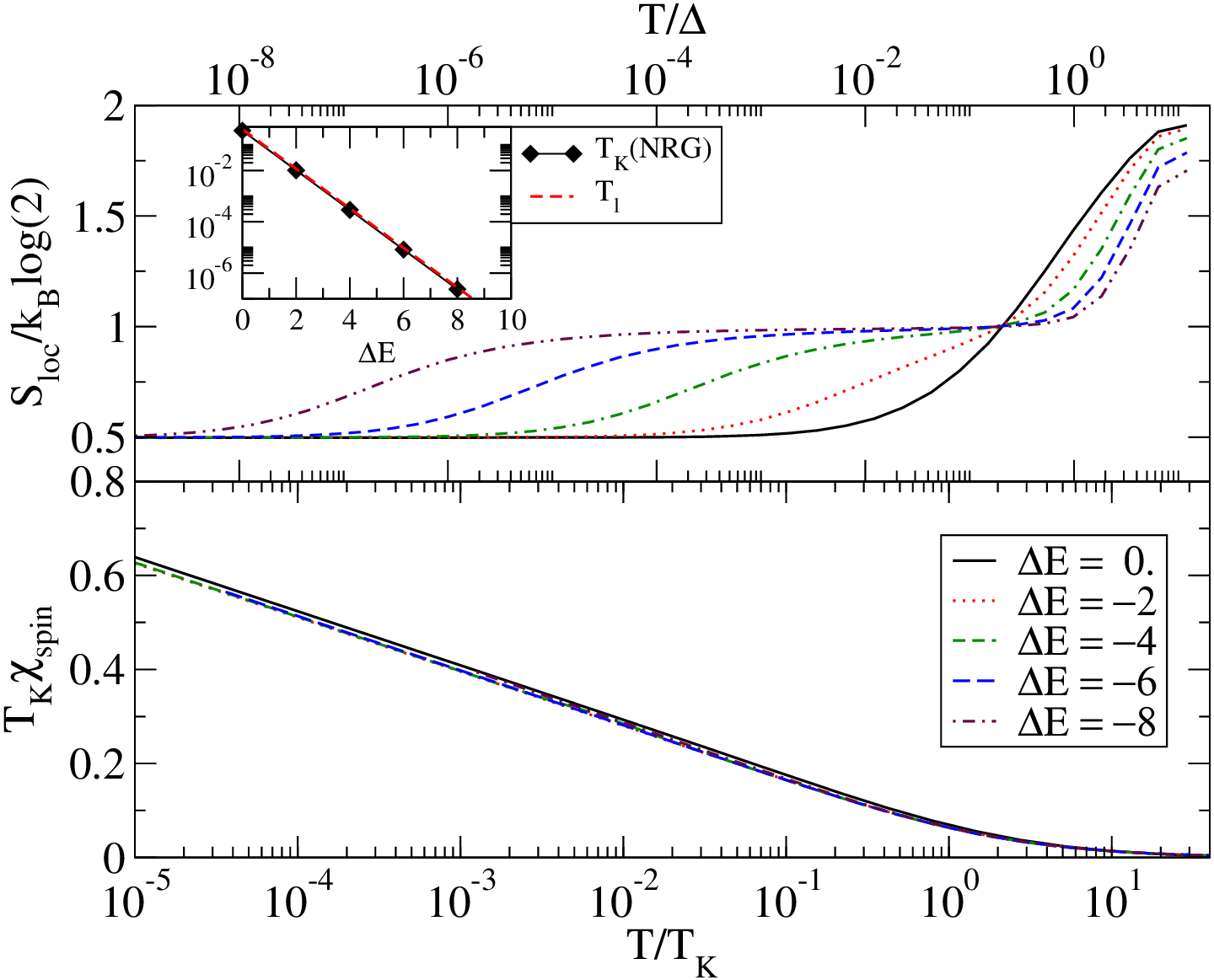}
  \caption{Entropy vs temperature for different values of $\Delta
    E=E_{\as}$. For $|\Delta E/\Delta|\ll 1$, the Hamiltonian flows
    directly from the free orbital fixed point $H^*_{FO}$
    ($S_{FO}=k_B\log(4)$)  to the two channel
    fixed-point ($S_{tc}= \log(2)/2$). For large values $|\Delta
    E/\Delta|\gg 1$, the Hamiltonian crosses over 
    to the local moment fixed point ($S_{LM}=\log(2)$) first,
    before reaching the line of two channel fixed-points $H_{tc}^*$. $T_K$
    associated with the quenching of the lower doublet is 
    shown in the inset. Parameters: $D=10$, $N_s=3000$, $\Lambda=3.5$.
}
  \label{fig:entropy-spin-sus}
\end{figure}

\paragraph{Thermodynamic results:}
A symmetric and constant band density $\rho_0=1/(2D)$, a band width of
$D=10\Delta$, and a NRG discretization parameter $\Lambda=3.5$ was chosen. All
energies are measured in units of the Anderson width $\Delta$. At each
NRG iteration, $N_s=3000$ states are kept. The impurity entropy $S_{imp}$
is defined as the difference of the entropy of (\ref{eqn:tcsiam-h}) with
or without the quantum impurity \cite{Wilson75}: $S_{imp} =
S_{full}-S_{free}$. It is plotted for different values of $|\Delta E|$
in the upper part of Fig.~\ref{fig:entropy-spin-sus} and agrees
excellently with the Bethe ansatz results\cite{BolechAndrei02}. For large
values of $|\Delta E/\Delta|\gg 1$, the reduction of the entropy from
$\log(4)\to\log(2)$ occurs on the scale  $T_h$, while the two-channel
fixed point is approached below 
$T_K$\cite{SchillerAndCox98,BolechAndrei02}. Even though the 
two-channel fixed point is always characterized by a residual entropy
of $\frac{1}{2}\log(2)$, the NRG level spectrum of the fixed-point is
non-universal and dependent on $\Delta E$ or the occupation number
$n_c$. Since the model lacks particle-hole 
symmetry away from $\Delta E=0$, the additional marginal operator,
$H_m=K(\sum_{\as} f^\dagger_{\as 0}f_{\as 0}-2)$
\cite{KrishWilWilson80ab}, leads to a line of fixed 
point Hamiltonians  comprising of $H^*_{tc}(\Delta E=0)$  and
$H_m$. The constant $K$ can be extracted from the NRG level spectra
and contains the $n_c$-dependence \cite{KrishWilWilson80ab}.

The logarithmically divergent susceptibilities to the low 
temperature form $\chi(T)= -\frac{1}{20T_K}\log(T) +b$ was fitted to extract
$T_K$. $T_K$ can also be defined as the temperature at which the
effective moment  of the lower doublet, $\mu^2_{eff}(T) =T\chi(T)$, is
reduced to $\mu_{eff}^2(T_K)=0.07$ \cite{KrishWilWilson80ab}. These two
temperatures coincide within the numerical error. The local spin susceptibility
$\chi_{s}$ for $\Delta E\le 0$, or channel susceptibility $\chi_{c}$ for
$\Delta E\ge 0$ is plotted in the lower graph of
Fig.~\ref{fig:entropy-spin-sus} for five different values of $\Delta
E$ as a function of $T/T_K$, to illustrate the scaling. The inset shows
the exponential behavior of $T_K(NRG)$ and the excellent agreement with the
Bethe ansatz wide band limit $T_K = \frac{4\Delta}{\pi^2} \exp[-
\frac{A_\lambda\pi |\Delta E|}{2\Delta }]$, taking into account
the discretization correction $A_\Lambda =  
\frac{1}{2}\frac{1+1/\Lambda}{1-1/\Lambda} \log \Lambda$ for the Kondo
coupling \cite{KrishWilWilson80ab}: $J\to J_{eff}=J/A_\Lambda$.

\begin{figure}[t]
  \centering
  \includegraphics[width=85mm,clip]{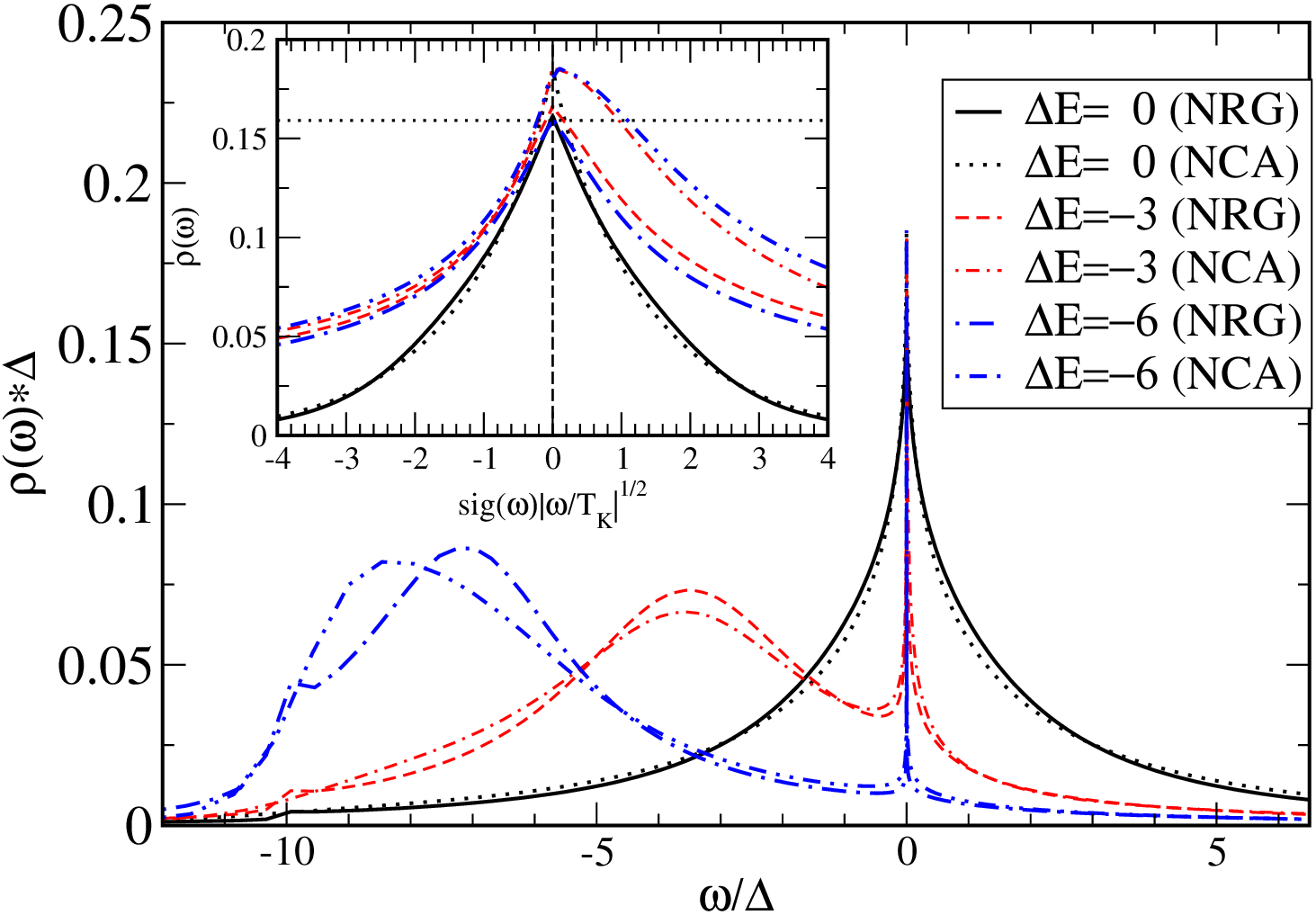}
  \caption{NRG spectral functions   for three different
    values of $\Delta E$  at $T=0$ in comparison with NCA spectral
    functions. The inset shows the behavior
    for $|\w|\le T_K$. The dotted lines marks $1/(2\pi\Delta)$. For
    increasing $|\Delta E|$, the peak of the NRG 
    spectrum becomes more symmetric and approaches a universal curve.
    The NCA spectra are calculated at $T=10^{-6}T_K$ for $\Delta
    E=0,3$ and  $T=10^{-3}T_K$. The value of $T_K$ are 
    $0.405285,3.6*10^{-3},3.3*10^{-5}$ for
    $\Delta E=0,-3,-6$.
   Parameters: $D=10$, $N_s=3000$, $\Lambda=2.2$.
}
  \label{fig:spectrum-nrg-nca-t0}
\end{figure}


\paragraph{Dynamical properties:}
Using the equation of motion for Fermionic  Green functions,
$z \green{A}{B} = \langle \{ A, B\}\rangle + \green{[A,H]}{B}$, the
commutators of the operators $X_{\alpha,\sigma}$ and $c_{k,\as}$ with
the Hamiltonian, 
$[X_{\alpha,\sigma}, H]$ and $[c_{k,\as},H]$ respectively, 
and the local completness $\sum_\sigma X_{\sigma\sigma} +\sum_\alpha
X_{\alpha\alpha} =1$, it is
straight forward to derive
\begin{eqnarray}
  [z- E_{\as} -\Gamma_{\as}]
    \green{X_{\alpha,\sigma}}{X_{\sigma,\alpha}}
&=& 
\nonumber \\
P_{\alpha\sigma}  +\green{\hat O_{\alpha\sigma}}{X_{\sigma,\alpha} }
\; .
\end{eqnarray}
The expectation value of anti-commutator
$P_{\as} =  \langle X_{\alpha,\alpha}\rangle + \langle
X_{\sigma,\sigma}\rangle$ is alway $1/2$ in the absence of a symmetry
breaking field. The local composite operator 
 $\hat O_{\alpha\sigma}$ 
\begin{eqnarray}
\hat O_{\alpha\sigma} 
&=&
\sum_k V_{-\alpha}
  \left(X_{-\sigma,\sigma}c_{k-\alpha-\sigma}
        -X_{-\sigma,-\sigma}c_{k-\alpha\sigma}
\right)
\nonumber \\
&&
+
\sum_k 
  \left(V_{\alpha} X_{\alpha,-\alpha}c_{k\alpha\sigma}
       -V_{-\alpha} X_{-\alpha,-\alpha}c_{k-\alpha\sigma}
\right)
\label{eqn:op-oas}
\end{eqnarray}
also transforms as $X_{\alpha,\sigma}$.
If  the Green function $G_{\as}(z)$ is parameterized by the  self-energy
$\Sigma_{\as}(z)$, $G_{\as}(z) = P_{\as}[z- E_{\as} -\Gamma_{\as}
-\Sigma_{\as}(z)]^{-1}$, this self-energy is given by the exact
relation
\begin{eqnarray}
  \Sigma_{\as}(z) &=&
M_{\as} (z)/G_{\as}(z)
\;\; ,
\label{sigma-nrg-tcsiam}
\end{eqnarray}
defining  $M_{\as} (z)= \green{O_{\alpha\sigma}}{X_{\sigma,\alpha}
}$. In contrary to  the single channel
model\cite{BullaHewsonPruschke98}, there exists  no non-interacting
limit, and  the spectral weight is reduced to  $P_{\as}=1/2$.

It is straight forward to calculate the matrix elements of operator
$\hat O_{\as}$  within the NRG. The spectral functions for $M_{\as}$
and  $G_{\as}$ are obtained from their Lehmann representation
by broading the $\delta$-function of the discret spectrum on a
logarithmic scale, i.~e.~$\delta(\w-E)\to e^{-b^2/4}
e^{-(\log(\w/E)/b)^2}/(\sqrt{\pi} b|E|)$
~\cite{BullaHewsonPruschke98}. The broadening  
parameter is usually chosen as $0.5 < b <1$, and we used $b=0.8$.
The self-energy $\Sigma(z)$ is then calculated through
(\ref{sigma-nrg-tcsiam}). Fig.~\ref{fig:spectrum-nrg-nca-t0} shows the
spectral function $\rho_{\as}(\w)$  in comparison with the
results obtain by the NCA\cite{Cox87}. The overall
agreement between the NRG and NCA curves for the same parameter is
reasonably good in spite of the incorrect threshold exponents of 
the NCA. The NRG tends to overestimate the broadening of the high
energy peaks at $\w\approx\Delta E$, but produces highly accurate
results for small frequencies $|\w|<T_K$. The inset of
Fig.~\ref{fig:spectrum-nrg-nca-t0} depicts the many-body resonance in
the region for $|\w/T_K|<10$ on a rescaled axis $|\w/T_K|^{1/2}$. 
It highlights  three major aspects of spectral function: (i) $\rho(\w)
\propto 1-a_\pm|\w/T_K|^{1/2}$ for $|\w|<T_K$, (ii) for increasing
$|\Delta E|$, the spectral function becomes more symmetric around
$\w=0$  and should approach a universal curve, (iii) the peak value is
pinned to $1/(2\pi\Delta)$ at $\w=0$ independent of $\Delta E$.
Even though the vertex correction to the NCA spectra  vanishs
only in the limit of large spin and channel degeneracy N and
M\cite{CoxRuck93}, the scaling behavior of the NCA spectra is
apparently correct. The NCA peak, however, saturates at
$\pi/(16\Delta)$ for $T\to 0$ \cite{CoxZawa98} and  remains asymmetric
for $|\Delta E|\gg \Delta$. 

\begin{figure}[t]
  \centering
  \includegraphics[width=85mm,clip]{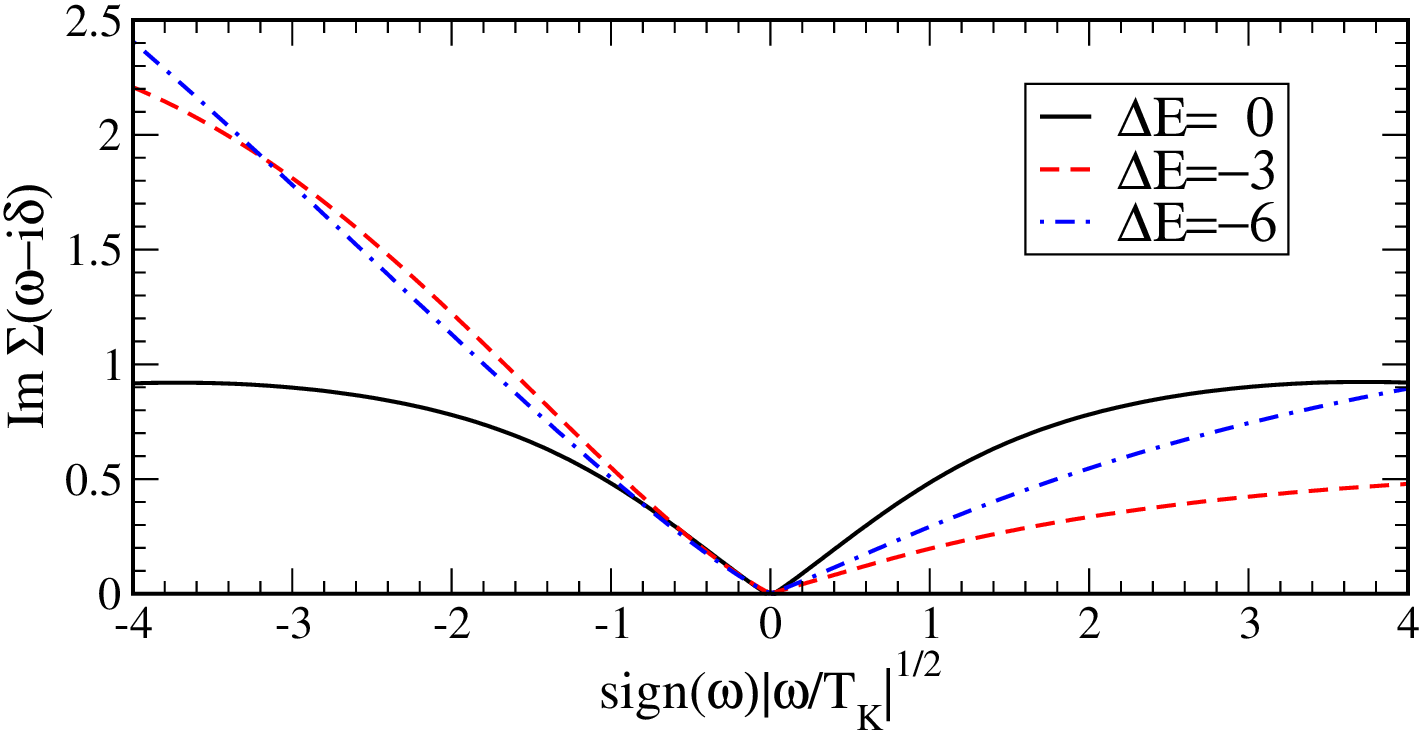}
  \caption{Imaginary part of the self-energy $\Sigma(\w-i\delta)$  as
    function of $\mbox{sign}(\w)\sqrt{|w/T_K|}$ for same three
    different  values of $\Delta E=E_{\as}$ at $T=0$ and parameters as
    Fig.~\ref{fig:spectrum-nrg-nca-t0}.
  }
  \label{fig:self-energy-nrg-nca-t0}
\end{figure}

Fig.~\ref{fig:self-energy-nrg-nca-t0} shows the corresponding
imaginary part of the NRG self-energy $\Sigma(\w-i\delta)$:  it vanishes at
$\w=0$, independent of $\Delta E$, which is a neccessary condition for the
pinning of the spectral  function for all values of $\Delta E$. In
addition, $\Re e \Sigma(0) +\Delta +E_{\as}=0$, leading to the
pinning of the resonance to $\w=0$, depicted in
Fig.~\ref{fig:spectrum-nrg-nca-t0}.   In the scaling regime
$|\w/T_K| < T_h$, $\Im m\Sigma(\w-i\delta) \propto
\sqrt{|\w/T_K|}$.  This region  extends to multiples of $|\w/T_K|$ for
$T_h\gg T_K$.   The definition of the self-energy via the
parameterization of $G_{\as}(z)$  turns out to be very useful for the
understanding of its properties. The  conformal field theory result for
the $T$-matrix of two-channel Kondo model \cite{Affleck91} conveys to
the TC-SIAM.  The self-energy $\Sigma(z)$ exhibits similar features as
the correlation self-energy $\Sigma^U(z)=\Sigma(z)-\Gamma(z)$ of the
single-channel model: its imaginary part also vanishes for $\w\to 0$
and its scaling behavior reflects the nature of the local electronic
excitation. While $\Im m\Sigma(\w-i\delta) \propto (\w/T_K)^2$ 
in the single channel model,  the scaling as $\Im m\Sigma(\w-i\delta) \propto\sqrt{|\w/T_K|}$ represents the
hallmark of a local non-Fermi liquid.

\begin{figure}[t]
  \centering
  \includegraphics[width=85mm,clip]{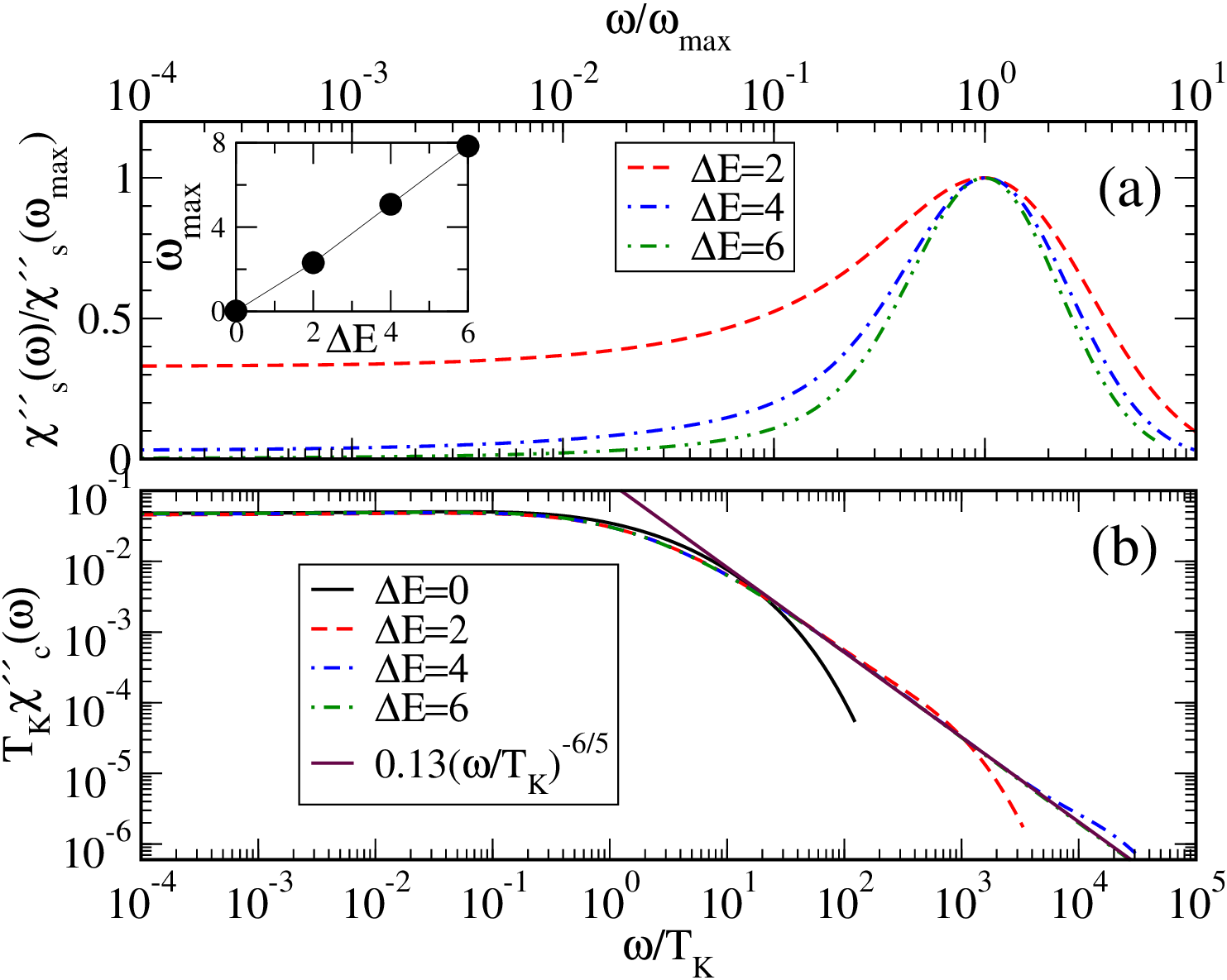}
  \caption{Imaginary part of $\chi''_{s}(\w)$ (a) scaled to position
    and value of the     finite frequency maximum  $\w_{max}\approx
    T_h$ as displayed 
    in the inset; the values of $\chi''_{s}(\w_{max})$ are
    $0.0126, 0.0046, 0.00235$ for $\Delta E = 2,4,6$. The
    scaling curve (b) for     $\chi''_{c}(\w)$ vs $\w/T_K$ at
    $T=0$ for $\Delta E \ge 0$.     Parameters as
    Fig.~\ref{fig:spectrum-nrg-nca-t0}.
  }
  \label{fig:dynamical-spin-sus-t0}
\end{figure}

The imaginary parts of the dynamical impurity  spin and quadrupolar or
charge \sus, $\chi_s(\w)=\green{\hat s_z}{\hat  s_z}$ and $\chi_c(\w)=
\green{\hat  \tau_z}{\hat \tau_z}$ respectively, are plotted in
Fig.~\ref{fig:dynamical-spin-sus-t0} for $T=0$ and positive $\Delta
E\ge 0$. Since the Hamiltonian is symmetric with respect to a
particle-hole transformation, and simmultanious $\Delta E\to -\Delta
E$ and interchanging the spin and quadrupolar indices,
Fig.~\ref{fig:dynamical-spin-sus-t0} also contains the information for
$\Delta E < 0$.  Both $\chi''$ show  two frequency regimes. In
contrast to the single channel model where the  relation $\chi''
\propto \w$ holds, $\chi''_{s/c}(\w)$  approachs a constant at low
frequencies, $|\w|<T_K$. This corresponds to the logarithmic
divergence of {\em   both} $\chi(T)$ for $T<T_K$, and  the slope of
the logarithm is given by this constant. A scaling curve is found for
the charge \sus\ (or spin \sus\ for $\Delta E<0$) when plotted in
dimensionless units of the low temperature scale $T_K$, as shown in
Fig.~\ref{fig:dynamical-spin-sus-t0} (b). The scaling holds for
$\w/T_h < 1$ above which non-universal band edge effects govern the
physics. The  intermediate frequency range $T_K < |\w |< T_h$ is
characterised by a power law behavior: $\chi''_{c}(\w) \propto
(\w/T_K)^{-1.2}$. The peak position in  $\chi''_{s}(\w)$ scales with
$T_h$ and corresponds to $\Delta E$ for large $\Delta E$
\cite{SchillerAndCox98,BolechAndrei02}.  The values of
$\chi''_{s}(\w_{max})$  become increasing smaller with reduction of
the occupance of the spin-doublet.


\paragraph{Discussion and Conclusion:}
The dynamical properties of the two-channel Anderson model were
accuratly  calculated for $T=0$ using the NRG approach. The analytical
form  of the  spectral function  is given by
$1/(2\pi\Delta)(1-a_\pm\sqrt{\w/T_K})$ for $|\w| < T_K$. Its peak is
pinned to a universal value, and 
the self-energy scales as $\sqrt{\w/T_K}$ independent of $n_c$.
Since the self-energy $\Sigma(z)$ does not
contain the resonant level part $\Delta_{\as}$, and $\Im
m\Sigma(\w-i\delta)\to 0$ for $\w\to 0$, its anomalous scaling serves
as a hallmark for local non-Fermi liquid behaviour. The NCA result for
the spectral function agrees remarkably well with the accurate
NRG result.  The two energy scales characterising the screening of the
moments of upper and lower doublets also appear as charge exitation energy
and as effective width of the many-body resonance at $\w=0$ in the
spectral function. At finite temperatures - not shown here - the
many-body resonance decreases for inceasing temperature and vanishes
for $T\gg T_K$. 
The dynamical \sus\ of the lower doublet maps to a universal scaling
curve for $\w < \mbox{min}\{D,T_h\}$ and $T=0$, while a peak at $\w_{max}\approx
T_h$ is found in the dynamical \sus\ of the upper doublet.

We have benefited greatly from fruitful discussions with  N.~Andrei,
C.~Bolech and A.~Schiller and would like to  acknowledge the  funding
of the NIC, Forschungszentrum  J\"ulich, under project no. HHB00.

%

\end{document}